\documentclass{article}

\usepackage{microtype}
\usepackage{graphicx}
\usepackage{subfigure}
\usepackage{booktabs} 
\usepackage{amsmath} 

\usepackage[frozencache]{minted}

\usepackage{todonotes}
\usepackage{colortbl}
\definecolor{deepblue}{rgb}{0,0,0.5}
\definecolor{deepred}{rgb}{0.6,0,0}
\definecolor{deepgreen}{rgb}{0,0.5,0}
\usepackage{arydshln}  
\usepackage{tikz}
\usetikzlibrary{calc}

\usepackage{hyperref}

\usepackage[accepted]{mlsys2020style/mlsys2020}

\mlsystitlerunning{Autobatching}

\begin{document}

\twocolumn[
\mlsystitle{Automatically batching control-intensive programs for modern accelerators}



\mlsyssetsymbol{equal}{*}

\begin{mlsysauthorlist}
\mlsysauthor{Alexey Radul}{goog}
\mlsysauthor{Brian Patton}{goog}
\mlsysauthor{Dougal Maclaurin}{goog}
\mlsysauthor{Matthew D.\ Hoffman}{nyc}
\mlsysauthor{Rif A.\ Saurous}{mtv}
\end{mlsysauthorlist}

\mlsysaffiliation{goog}{Google, Cambridge, Massachusetts, USA}
\mlsysaffiliation{nyc}{Google, New York, New York, USA}
\mlsysaffiliation{mtv}{Google, Mountain View, California, USA}

\mlsyscorrespondingauthor{Alexey Radul}{axch@google.com}

\mlsyskeywords{Automatic vectorization, Machine Learning, MLSys}

\vskip 0.3in

\begin{abstract}
We present a general approach to batching arbitrary computations for
accelerators such as GPUs.  We show orders-of-magnitude speedups using our method
on the No U-Turn Sampler
(NUTS), a workhorse algorithm in Bayesian statistics.  The central
challenge of batching NUTS and other Markov chain Monte Carlo
algorithms is data-dependent control flow and recursion.  We overcome
this by mechanically transforming a single-example implementation into
a form that explicitly tracks the current program point for each batch
member, and only steps forward those in the same place.  We present
two different batching algorithms: a simpler, previously published one
that inherits recursion from the host Python, and a more complex,
novel one that implemenents recursion directly and can batch across it.
We implement these batching methods as a general program
transformation on Python source.  Both the batching system and the
NUTS implementation presented here are available as part of the
popular TensorFlow Probability software package.
\end{abstract}
]



\printAffiliationsAndNotice{}  

\section{Introduction}
\label{introduction}

Modern machine learning accelerators such as GPUs are oriented
around Single Instruction Multiple Data (SIMD) parallelism---doing
the same thing to each item of a big array of data at once.  Machine
learning programs optimized for such accelerators generally consist of
invoking \emph{kernels}, where each kernel is a separately hand-tuned
accelerator program for a specific function.  Good
utilization of the accelerator comes of making
relatively few kernel calls, with each
kernel processing a relatively large amount of data.  In the case of a
typical neural network workload, the kernels would be ``matrix
multiplication'' or ``convolution'', and the call sequence would
encode the architecture of the neural network.

Let's briefly look at the resulting programming model.  This review is
worded in the TensorFlow \cite{tensorflow2015-whitepaper} ecosystem,
since that's the setting for our work, but other machine learning
frameworks are broadly similar.  The top-level program is generally
written in Python, calling TensorFlow API functions that correspond to
kernels such as matrix multiplication.  These functions can be
executed immediately, in the so-called TensorFlow \emph{Eager mode}.
In this case they can be arbitrarily interleaved with the host Python,
including control flow; but suffer corresponding dispatch and
communication overhead.  Alternately, the same API functions can be
used to construct an operation graph to be executed all at once.  This
is the so-called TensorFlow \emph{graph mode}.  The advantage is that
graphs can be saved, loaded, and optimized before being run, and
suffer less dispatch overhead.  The disadvantage is that graph
computations cannot be interleaved with the host Python, and in
particular graph mode cannot represent recursive computations.  A
third option is to further compile the graph with XLA \cite{xla}.  XLA
imposes even more restrictions, such as statically resolving the
shapes of all intermediate arrays, but offers the additional benefit
of fusing kernels together, which reduces dispatch overhead even more.

Good performance in this programming style depends heavily on
vectorization, both within the kernels and at the level of kernel inputs.  One
very common strategy for vectorizing machine learning programs is
so-called \emph{batching}: processing a batch of
independent inputs in lock-step in order to get more play for
vectorization.  Batching can also reduce per-input memory pressure: in
the case of a neural network with $N$ features, each input has size
$O(N)$, whereas the weight matrices can easily have size $O(N^2)$.
Running multiple inputs through the layers of the network in lock-step
can re-use each weight matrix for many examples before having to evict
it from memory caches in order to load the next one.

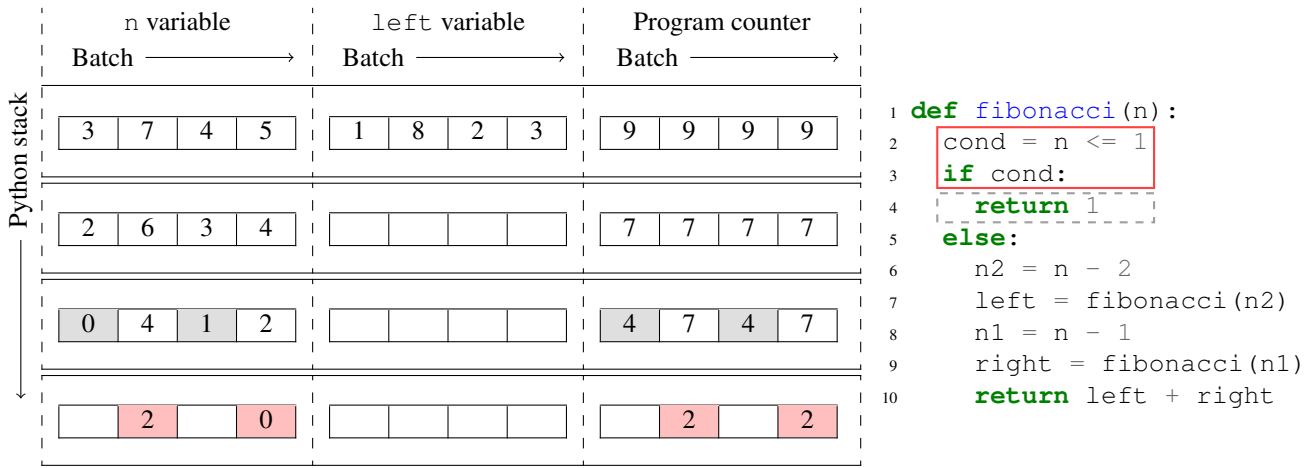
\begin{figure*}[ht]
  \noindent
\begin{minipage}[t]{0.03\textwidth}
\begin{tikzpicture}[remember picture,overlay]
  \node[rotate=90] (stack direction) at ($(current page.south west) + (65pt, 670pt)$) {
    Python stack
  };
  \draw[-to,shorten >=-1pt] ($(stack direction.west) + (0, -1pt)$) -- ($(stack direction.west) + (0, -60pt)$);

  \draw [draw=red!75,thick] (356pt,21pt) rectangle ++(82pt,23pt);
  \draw [draw=gray!75,thick,dashed] (356pt,8pt) rectangle ++(82pt,11pt);
\end{tikzpicture}
\end{minipage}
\begin{minipage}[t]{0.669\textwidth}
{\newlength{\colwidth}
\setlength{\colwidth}{1em}
\newlength{\outerrowheight}
\setlength{\outerrowheight}{2em}
\begin{tabular}{:c:c:c:}
  {\tt n} variable & {\tt left} variable & Program counter
  \\
  \begin{tikzpicture}
    \node (batch) at ($(0, 0)$) {
      Batch
    };

    \draw[-to,shorten >=-1pt] ($(batch.east) + (1pt, 0)$) -- ($(batch.east) + (56pt, 0)$);
  \end{tikzpicture}
  &
  \begin{tikzpicture}
    \node (batch) at ($(0, 0)$) {
      Batch
    };

    \draw[-to,shorten >=-1pt] ($(batch.east) + (1pt, 0)$) -- ($(batch.east) + (56pt, 0)$);
  \end{tikzpicture}
  &
  \begin{tikzpicture}
    \node (batch) at ($(0, 0)$) {
      Batch
    };

    \draw[-to,shorten >=-1pt] ($(batch.east) + (1pt, 0)$) -- ($(batch.east) + (56pt, 0)$);
  \end{tikzpicture}
  \\ \hline
  & & \\
  \noindent
  \begin{tabular}{|c|c|c|c|}
    \rule{\colwidth}{0pt}&\rule{\colwidth}{0pt}&\rule{\colwidth}{0pt}&\rule{\colwidth}{0pt}\\[-2.8ex]
    \hline
    3  & 7  & 4  & 5 \\ \hline
  \end{tabular}
  &
  \begin{tabular}{|c|c|c|c|}
    \rule{\colwidth}{0pt}&\rule{\colwidth}{0pt}&\rule{\colwidth}{0pt}&\rule{\colwidth}{0pt}\\[-2.8ex]
    \hline
    1  & 8  & 2  & 3 \\ \hline
  \end{tabular}
  &
  \begin{tabular}{|c|c|c|c|}
    \rule{\colwidth}{0pt}&\rule{\colwidth}{0pt}&\rule{\colwidth}{0pt}&\rule{\colwidth}{0pt}\\[-2.8ex]
    \hline
    9  & 9  & 9  & 9 \\ \hline
  \end{tabular}
  \\[1em] \hline \hline
  \rule{0pt}{\outerrowheight}%
  \noindent
  \begin{tabular}{|c|c|c|c|}
    \rule{\colwidth}{0pt}&\rule{\colwidth}{0pt}&\rule{\colwidth}{0pt}&\rule{\colwidth}{0pt}\\[-2.8ex]
    \hline
    2  & 6  & 3  & 4 \\ \hline
  \end{tabular}
  &
  \begin{tabular}{|c|c|c|c|}
    \rule{\colwidth}{0pt}&\rule{\colwidth}{0pt}&\rule{\colwidth}{0pt}&\rule{\colwidth}{0pt}\\[-2.8ex]
    \hline
       &    &    &   \\ \hline
  \end{tabular}
  &
  \begin{tabular}{|c|c|c|c|}
    \rule{\colwidth}{0pt}&\rule{\colwidth}{0pt}&\rule{\colwidth}{0pt}&\rule{\colwidth}{0pt}\\[-2.8ex]
    \hline
    7  & 7  & 7  & 7 \\ \hline
  \end{tabular}
  \\[1em] \hline \hline
  \rule{0pt}{\outerrowheight}%
  \noindent
  \begin{tabular}{|c|c|c|c|}
    \rule{\colwidth}{0pt}&\rule{\colwidth}{0pt}&\rule{\colwidth}{0pt}&\rule{\colwidth}{0pt}\\[-2.8ex]
    \hline
    \cellcolor{gray!25}0  & 4  & \cellcolor{gray!25}1  & 2 \\ \hline
  \end{tabular}
  &
  \begin{tabular}{|c|c|c|c|}
    \rule{\colwidth}{0pt}&\rule{\colwidth}{0pt}&\rule{\colwidth}{0pt}&\rule{\colwidth}{0pt}\\[-2.8ex]
    \hline
       &    &    &   \\ \hline
  \end{tabular}
  &
  \begin{tabular}{|c|c|c|c|}
    \rule{\colwidth}{0pt}&\rule{\colwidth}{0pt}&\rule{\colwidth}{0pt}&\rule{\colwidth}{0pt}\\[-2.8ex]
    \hline
    \cellcolor{gray!25}4  & 7  & \cellcolor{gray!25}4  & 7 \\ \hline
  \end{tabular}
  \\[1em] \hline \hline
  \rule{0pt}{\outerrowheight}%
  \noindent
  \begin{tabular}{|c|c|c|c|}
    \rule{\colwidth}{0pt}&\rule{\colwidth}{0pt}&\rule{\colwidth}{0pt}&\rule{\colwidth}{0pt}\\[-2.8ex]
    \hline
       & \cellcolor{red!25}2  &    & \cellcolor{red!25}0 \\ \hline
  \end{tabular}
  &
  \begin{tabular}{|c|c|c|c|}
    \rule{\colwidth}{0pt}&\rule{\colwidth}{0pt}&\rule{\colwidth}{0pt}&\rule{\colwidth}{0pt}\\[-2.8ex]
    \hline
       &    &    &   \\ \hline
  \end{tabular}
  &
  \begin{tabular}{|c|c|c|c|}
    \rule{\colwidth}{0pt}&\rule{\colwidth}{0pt}&\rule{\colwidth}{0pt}&\rule{\colwidth}{0pt}\\[-2.8ex]
    \hline
       & \cellcolor{red!25}2  &    & \cellcolor{red!25}2 \\ \hline
  \end{tabular}
  \\[1em] \hline
\end{tabular}}
\end{minipage}
\begin{minipage}[t]{0.28\textwidth}
\vspace{-0.75in}
\begin{minted}[
    mathescape,
    linenos,
    numbersep=5pt,
    frame=none,
    framesep=2mm]{python}
def fibonacci(n):
  cond = n <= 1
  if cond:
    return 1
  else:
    n2 = n - 2
    left = fibonacci(n2)
    n1 = n - 1
    right = fibonacci(n1)
    return left + right
\end{minted}
\end{minipage}

  \caption{Runtime operation of a locally, statically auto-batched recursive
    Fibonacci program.  This snapshot occurs on the batch of inputs $3, 7, 4, 5$.
    The batching transformation adds storage for
    all the batch members and handles divergent control flow by
    masking.  The recursion is handled in Python.  In this example,
    there are two ``active'' logical threads about to execute lines
    2-3 of the Fibonacci program, highlighted in red.  There are also
    two logical threads suspended one Python stack frame earlier,
    waiting for the active threads to re-converge with them so they
    can all return from that frame.  The runtime cannot batch together
    logical threads with different call stacks, because those stacks
    are embedded in the runtime's Python-level call stack.  The \texttt{left}
    variable has no value in most of the shown stack frames because the
    program hasn't assigned it yet.}
  \label{fig:python-stack-idea}
\end{figure*}

It is standard practice in machine learning frameworks
such as TensorFlow or PyTorch \cite{paszke2017automatic} to code the
kernels to accept extra input dimensions and operate elementwise across them.  Consequently,
coding a batched version of a straightline program is relatively
straightforward, if somewhat tedious and error-prone.  Simple
neural networks being straightline, batch
training is the norm.  Obstacles arise, however, when one wishes to
batch a program with control flow, such as conditionals or
variable-length loops.  Then it becomes necessary to keep track of
which batch member takes which branch of each conditional, and avoid
or ignore computations on batch members at the wrong program point.
The difficulty
of doing this by hand impedes using
sophisticated classical algorithms in machine learning.  Despite the
impedance, people have used tree searches \cite{alphago-paper-nature-2016},
optimization routines \cite{amos2017optnet} and
ordinary differential equations solvers \cite{neural-odes-nips-2018} in
machine learning work; what else could we accomplish if it were
easier?

Additional obstacles arise when trying to run a recursive program on a
modern machine learning framework, in batch or otherwise, because the
dataflow graph representation cannot execute recursion natively.
This is as true in XLA or TensorFlow graph mode as it is in other
graph-oriented machine learning frameworks like Caffe
\cite{jia2014caffe}.  The user is therefore forced to fall back to
eager-style execution, paying more
communication overhead.  If machine learning is to benefit fully from
the last 60 years of computer algorithm development, we must be able
to run recursive algorithms reasonably efficiently.

Our goal in this paper is to push the boundary of what classical
algorithms can efficiently execute on accelerators, in the context of
modern machine learning frameworks.  In particular, we
\begin{itemize}
\item Introduce \emph{program-counter autobatching}
  (Section~\ref{sec:tensor-stack}), a global, static program
  transformation for batching programs with arbitrary control flow,
  and materializing recursion into an underlying dataflow system.
\item Demonstrate that program-counter autobatching can successfully
  accelerate the No U-Turn Sampler, a classic algorithm from Bayesian
  statistics, by compiling its recursion into explicit stack
  management, and by statically constructing a schedule for running it
  on batches of inputs.
\item Provide, using the same vocabulary, a formal description of
  \emph{local static autobatching} (Section~\ref{sec:python-stack}).
  This is a simpler and lower-overhead batching transformation
  with less batching power in the recursive case.
\item Survey (Section~\ref{sec:related-work}) the local static
  autobatching systems \cite{pfor-icml-2019, matchbox-systems-for-ml-workshop-2018, jax} that have been implemented for several machine
  learning frameworks.
\item Directly compare these two autobatching strategies on a test
  problem from Bayesian statistics (Section~\ref{sec:results}).
\end{itemize}

Program-counter autobatching is available as a module in the popular
TensorFlow Probability \cite{tfp-repo, dillon2017tensorflow} software package.
That module also implements
a local static autobatching variant for comparison.

\section{Local Static Autobatching}
\label{sec:python-stack}

The simplest batching strategy (whether automated or hand-coded) is to
retain the graph of the computation as-is and just transform every
operation into a batched equivalent.  We call this \emph{local static
  autobatching}.  Intuitively, it's ``local'' because the pattern of
composition of operations doesn't change, and every operation can be
transformed on its own; and it's ``static'' because the batching
schedule doesn't depend on the input data, and can thus be computed
before starting execution.

When extending this idea to programs with control flow, it is
necessary to at least introduce a mask of which batch members are
``currently active''.  One then arranges to execute every control path
that at least one batch member follows, and avoid or ignore each
computation for each batch member that did not take that path.  If the
program being batched is recursive, the recursion still has to be
carried out by the control language, i.e., Python.  The runtime
operation thus looks like Figure~\ref{fig:python-stack-idea}.

Local static autobatching can be implemented in many styles.  For the
sake of clarity, we will describe it as a nonstandard interpretation
of a simple control flow graph language, given in
Figure~\ref{fig:python-stack-syntax}.  In addition to eliminating many
incidental considerations, this presentation aligns with the
presentation of program-counter autobatching in
Section~\ref{sec:tensor-stack}, which will be a (different)
nonstandard interpretation of a very similar language.  Going through
this presentation first also allows us to compare to other local
static autobatching systems more precisely, in
Section~\ref{sec:related-work}.

\begin{figure}[t]
$\arraycolsep=4pt
\begin{array}{rlll}
  \mathrm{Program}    & P  & ::= & [F] \\
  \mathrm{Function}   & F  & ::= & \mathrm{input}\ x, \mathrm{body}\ [B], \mathrm{output}\ y \\
  \mathrm{Block}      & B  & ::= & [op], t \\
  \mathrm{Operation}  & op & ::= & \mathrm{Primitive}\ y = f(x)  \\
                      &    &     & ~\vert~ \mathrm{Call}\ y = F(x) \\
  \mathrm{Terminator} & t  & ::= & \mathrm{Jump}\ i ~\vert~ \mathrm{Branch}\ x\ i\ j ~\vert~ \mathrm{Return} \\
  & f & ::= & \sin ~\vert~ \cos ~\vert~ \ldots
\end{array}$
\caption{Syntax of locally batchable programs.  We use $[\cdot]$ to
  denote ordered lists. The symbols $x$, $y$ range over variable names, and
  $i$, $j$ index blocks within the same function.  We present a unary
  syntax for succinctness; the $n$-ary generalization is standard.}
\label{fig:python-stack-syntax}
\end{figure}

The nonstandard interpretation itself is given in
Algorithm~\ref{alg:python-stack}.  In addition to storage for all the
batch member inputs, we maintain an \emph{active set} (initially the
whole batch) and a \emph{program counter} (initially the start of the
entry point).  The active set is a mask---all inactive batch members
are ignored and never modified until they become active.  The program
counter gives the program point (as a basic block index) each active
batch member is waiting to execute.  The execution model is simple: at
each step, we select some basic block that has at least one active
batch member and execute it in batch.  We then update the data storage
and program counters of just those \emph{locally active} batch members.  Repeat
until all active batch members have exited the function, then return.

\begin{algorithm}[tb]
   \caption{Local static autobatching}
   \label{alg:python-stack}
\begin{algorithmic}
   \STATE {\bfseries Input:} Function $F$ with $I$ basic blocks $B_i$, input variable $x$, and output variable $y$;
   \STATE {\bfseries Input:} Batch size $Z$;
   \STATE {\bfseries Input:} Data array $T$ with leading dimension $Z$;
   \STATE {\bfseries Input:} Active set $A \subseteq \{0, 1, \ldots, Z-1\}$.
   \STATE Initialize length $Z$ program counter $pc = [0, 0, \ldots, 0]$
   \STATE Initialize $x = T$
   \WHILE{(for any $b \in A$, $pc_b < I$)}
   \STATE Set block index $i = \min_{b \in A}{pc_b}$
   \STATE Compute locally active set $A' = \{b \in A | pc_b = i\}$
   \FOR{$op \in B_i$}
   \IF{$op$ is $(\mathrm{Primitive}\ y = f(x))$}
   \STATE Compute outputs $o = f(x)$
   \STATE Set $y_{A'} = o_{A'}$
   \ELSIF{$op$ is $(\mathrm{Call}\ y = G(x))$}
   \STATE Recursively compute outputs:
   \STATE $o = \textrm{Local-static}(G, Z, x, A')$
   \STATE Set $y_{A'} = o_{A'}$
   \ENDIF
   \ENDFOR
   \IF{$t_i$ is $\mathrm{Jump}\ j$}
   \STATE Set $pc_{A'} = j$
   \ELSIF{$t_i$ is $\mathrm{Branch}\ x\ j\ k$}
   \FOR{$b \in A'$}
   \STATE Set $pc_b = j$ if $x_b$ otherwise $pc_b = k$
   \ENDFOR
   \ELSIF{$t_i$ is $\mathrm{Return}$}
   \STATE Set $pc_{A'} = I$
   \ENDIF
   \ENDWHILE
   \STATE \textbf{return} Current value of $y$
\end{algorithmic}
\end{algorithm}

If the block we are executing ends in a branch (i.e., the prelude of a
source language \texttt{if} statement), the locally active batch members may
\emph{diverge}, in that some may move to the true branch and some to
the false.  They will \emph{converge} again when both of those
branches complete, and we continue after the end of the \texttt{if}.

If the block we are executing contains a (potentially recursive) call
to a function the user asked us to auto-batch, we appeal to the host
language's function call facility.  The only trick is to update the
active set in the recursive autobatching invocation to include only
the locally active batch members (i.e., those whose program counter
was at that call).

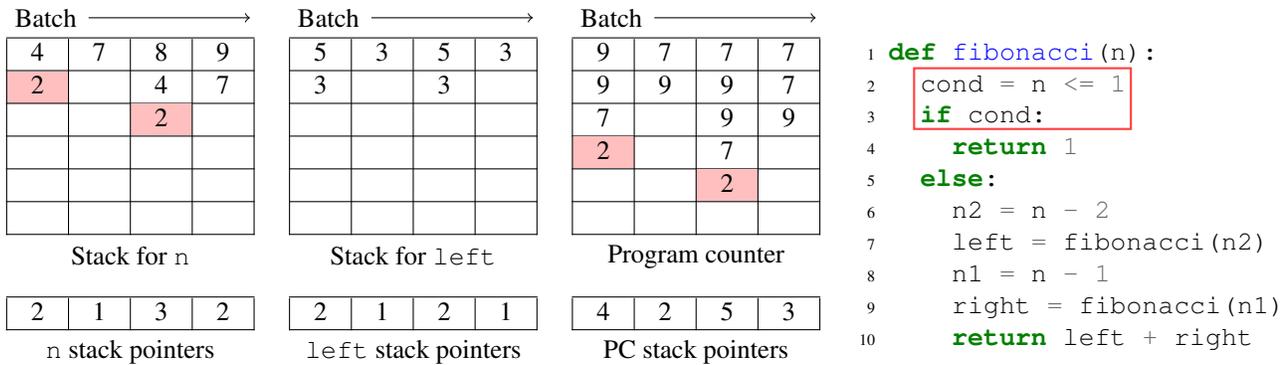
\begin{figure*}[ht]
  \noindent
\begin{minipage}[t]{0.679\textwidth}
\setlength{\colwidth}{1.1em}
\newlength{\smcolwidth}
\setlength{\smcolwidth}{\colwidth}
\begin{tikzpicture}
  \node[inner sep=0, outer sep=0](var n) at (0, 0) {
    \begin{tabular}{|c|c|c|c|}
      \rule{\smcolwidth}{0pt}&\rule{\smcolwidth}{0pt}&\rule{\smcolwidth}{0pt}&\rule{\smcolwidth}{0pt}\\[-2.8ex]
      \hline
      4                   & 7 & 8                   & 9 \\ \hline
      \cellcolor{red!25}2 &   & 4                   & 7 \\ \hline
                          &   & \cellcolor{red!25}2 &   \\ \hline
      & & & \\ \hline
      & & & \\ \hline
      & & & \\ \hline
    \end{tabular}};

  \node[inner sep=0, outer sep=0](var n pointers) at ($(var n.south) + (0, -30pt)$) {
    \begin{tabular}{|c|c|c|c|}
      \rule{\smcolwidth}{0pt}&\rule{\smcolwidth}{0pt}&\rule{\smcolwidth}{0pt}&\rule{\smcolwidth}{0pt}\\[-2.8ex]
      \hline
      2 & 1 & 3 & 2 \\ \hline
    \end{tabular}};

  \node (batch 2) at ($(var n.north west) + (1.5em, 8pt)$) {
    Batch
  };

  \draw[-to,shorten >=-1pt] ($(batch 2.east) + (1pt, 0)$) -- ($(batch 2.east) + (61pt, 0)$);

  \node at ($(var n.south) + (0, -8pt)$) {
    \centering
    Stack for \tt n
  };

  \node at ($(var n pointers.south) + (0, -8pt)$) {
    \centering
    \tt n\rm\ stack pointers
  };

  \node[inner sep=0, outer sep=0](var left) at ($(var n.east) + (60pt, 0)$) {
    \begin{tabular}{|c|c|c|c|}
      \rule{\smcolwidth}{0pt}&\rule{\smcolwidth}{0pt}&\rule{\smcolwidth}{0pt}&\rule{\smcolwidth}{0pt}\\[-2.8ex]
      \hline
      5 & 3 & 5 & 3 \\ \hline
      3 &   & 3 &   \\ \hline
      & & & \\ \hline
      & & & \\ \hline
      & & & \\ \hline
      & & & \\ \hline
    \end{tabular}};

  \node[inner sep=0, outer sep=0](var left pointers) at ($(var left.south) + (0, -30pt)$) {
    \begin{tabular}{|c|c|c|c|}
      \rule{\smcolwidth}{0pt}&\rule{\smcolwidth}{0pt}&\rule{\smcolwidth}{0pt}&\rule{\smcolwidth}{0pt}\\[-2.8ex]
      \hline
      2 & 1 & 2 & 1 \\ \hline
    \end{tabular}};

  \node (batch 3) at ($(var left.north west) + (1.5em, 8pt)$) {
    Batch
  };

  \draw[-to,shorten >=-1pt] ($(batch 3.east) + (1pt, 0)$) -- ($(batch 3.east) + (61pt, 0)$);

  \node at ($(var left.south) + (0, -8pt)$) {
    \centering
    Stack for \tt left
  };

  \node at ($(var left pointers.south) + (0, -8pt)$) {
    \centering
    \tt left\rm\ stack pointers
  };

  \node[inner sep=0, outer sep=0](program counter) at ($(var left.east) + (60pt, 0)$) {
    \begin{tabular}{|c|c|c|c|}
      \rule{\colwidth}{0pt}&\rule{\colwidth}{0pt}&\rule{\colwidth}{0pt}&\rule{\colwidth}{0pt}\\[-2.8ex]
      \hline
      9                   & 7 & 7                   & 7 \\ \hline
      9                   & 9 & 9                   & 7 \\ \hline
      7                   &   & 9                   & 9 \\ \hline
      \cellcolor{red!25}2 &   & 7                   &   \\ \hline
                          &   & \cellcolor{red!25}2 &   \\ \hline
                          &   &                     &   \\ \hline
    \end{tabular}};

  \node[inner sep=0, outer sep=0](program counter pointers) at ($(program counter.south) + (0, -30pt)$) {
    \begin{tabular}{|c|c|c|c|}
      \rule{\colwidth}{0pt}&\rule{\colwidth}{0pt}&\rule{\colwidth}{0pt}&\rule{\colwidth}{0pt}\\[-2.8ex]
      \hline
      4  & 2  & 5  & 3 \\ \hline
    \end{tabular}};

  \node (batch) at ($(program counter.north west) + (1.5em, 8pt)$) {
    Batch
  };

  \draw[-to,shorten >=-1pt] ($(batch.east) + (1pt, 0)$) -- ($(batch.east) + (61pt, 0)$);

  \node at ($(program counter.south) + (0, -8pt)$) {
    \centering
    Program counter
  };

  \node at ($(program counter pointers.south) + (0, -8pt)$) {
    \centering
    PC stack pointers
  };

  \draw [draw=red!75,thick] (0.895\textwidth,3pt) rectangle ++(82pt,23pt);

\end{tikzpicture}
\end{minipage}
\begin{minipage}[t]{0.3\textwidth}
\vspace{-1.73in}
\begin{minted}[
    mathescape,
    linenos,
    numbersep=5pt,
    frame=none,
    framesep=2mm]{python}
def fibonacci(n):
  cond = n <= 1
  if cond:
    return 1
  else:
    n2 = n - 2
    left = fibonacci(n2)
    n1 = n - 1
    right = fibonacci(n1)
    return left + right
\end{minted}
\end{minipage}
  \caption{Runtime operation of a program counter auto-batched
    recursive Fibonacci program.  This snapshot occurs on the batch of inputs $6, 7, 8, 9$.
    In addition to the batch dimension
    (across), the batching transformation also augments every
    non-temporary variable from the program with a stack dimension
    (down), and an array of stack pointers.  Additionally, the runtime
    maintains a \texttt{program counter} variable that records which
    block each logical thread is waiting to execute.  At each time
    step, the runtime selects a basic block to run (lines 2-3 in this
    example) and updates the state and program counter of the logical
    threads executing that block (highlighted in red).  Because
    recursive state is captured explicitly in the arrays storing the
    data, the runtime doesn't need to itself rely on recursion in
    Python (the host language).  This means both that it can be
    executed in TensorFlow's graph mode, and that it can let logical
    threads re-converge on function calls, even at different stack
    depths.  Note that the stack for the \texttt{n} variable will only
    hold values in frames where the program counter hasn't moved past
    line 8, where \texttt{n} is last used.  Conversely, \texttt{left}
    is only pushed in frames where the program counter is past line
    7.}
  \label{fig:tensor-stack-idea}
\end{figure*}

Why does this work?  Consider this runtime from the point of view of
one batch member.  It wants to execute some sequence of basic blocks,
as given by the edits to its program counter.  Every time the runtime
runs one of those basic blocks, it updates the state of that batch
member the same way it would if the batch had size 1.  And every time
the runtime runs some other block, it doesn't update the batch member
at all.  The only way this can fail is if some underlying batch
operation in the platform doesn't treat batch members independently
(e.g., if an error in one batch member causes an exception which
aborts execution of all of them) or if some batch member doesn't
terminate and starves the others.

There are two significant free choices in this runtime.  The first is
how to execute a primitive operation on some batch members but not
others.  Algorithm~\ref{alg:python-stack} is written in \emph{masking}
style: we run the primitive on all the batch members, and just ignore
the results of the ones that were at different points in the program.
This is simple and has very low in-system overhead, because masking is
a cheap operation.  The down side is that it wastes computation on the
batch members that are going to be masked out, which can be
significant if batch utilization is low.  There is also the subtlely
that this extra computation happens with junk data, which may trigger
spurious failures in the underlying platform.

The other option for batching the primitive operations is to use the
indices of the locally active batch members to gather the inputs into
a smaller array, perform just the live computation, and then scatter
the results back into the full runtime state.  This avoids wasting
computation and avoids computing on junk data, but gathering and
scattering are more expensive than masking.  Furthermore, the
intermediate arrays will have statically indeterminate size, making
the gather-scatter approach less effective on platforms like the XLA
compiler for Google's TPUs \cite{xla} that statically infer array
shapes.

The second significant free choice in this runtime is the heuristic
for selecting which basic block to run next.  As long as we don't
starve any blocks, any selection criterion will lead to a correct end
result.  Algorithm~\ref{alg:python-stack} encodes a surprisingly effective
choice: always run the earliest available block in program order.
This has the merit of being (relatively) predictable by the user; but
more refined heuristics are definitely possible.

In our implementation, the frontend for this is a Python-embedded compiler.
That is, it's a user-invoked AST
transformation based on AutoGraph \cite{moldovan2018autograph} that converts the user program into the form given in
Figure~\ref{fig:python-stack-syntax}.  All the user's actual
computations become $\mathrm{Primitive}$ operations, and the control
and recursion constructs are encoded in a standard way in
$\mathrm{Jump}, \mathrm{Branch}, \mathrm{Call}$, and $\mathrm{Return}$
instructions.

\section{Program Counter Autobatching}
\label{sec:tensor-stack}

The local static autobatching discussed in
Section~\ref{sec:python-stack} has an interesting limitation.  Because
it relies on the Python stack to implement recursion, it cannot batch
operations across different (recursive) calls to the same user
function.  So two batch members could be trying to execute the same
code and not be batchable because the system doesn't have a
long-enough sight line to connect them.  And, of course, relying on
Python to manage the recursion imposes communication costs and limits
the optimizations the underlying machine learning framework can do.

We can serve two purposes with one intervention by implementing the
stack within the autobatching system.  We choose to give each program
variable its own stack (by extending the relevant array with another
dimension), getting a runtime state that looks like
Figure~\ref{fig:tensor-stack-idea}.  The layout of the figure is intentionally
the same as Figure~\ref{fig:python-stack-idea} to
emphasize that we are representing all the same stuff, just in a
different way.

\begin{figure}[t]
$\arraycolsep=4pt
\begin{array}{rlll}
  \mathrm{Program}    & P  & ::= & \mathrm{input}\ x, \mathrm{code}\ [B], \mathrm{output}\ y \\
  \mathrm{Block}      & B  & ::= & [op], t \\
  \mathrm{Operation}  & op & ::= & \mathrm{Push}\ y = f(x) ~\vert~ \mathrm{Pop}\ x \\
  \mathrm{Terminator} & t  & ::= & \mathrm{Jump}\ i ~\vert~ \mathrm{Branch}\ x\ i\ j \\
                      &    &     & ~\vert~ \mathrm{PushJump}\ i\ j ~\vert~ \mathrm{Return} \\
  & f & ::= & \sin ~\vert~ \cos ~\vert~ \ldots \\
\end{array}$
\caption{Syntax of program counter batchable programs.  We use $[\cdot]$
  to denote ordered lists. The symbols $x$, $y$ range over variable names,
  and $i$, $j$ index blocks of the program.  This syntax is also unary for succinctness.
  The difference from
  locally autobatched programs (Figure~\ref{fig:python-stack-syntax}) is that
  all control flow graphs are merged, and $\mathrm{Call}$ operations are replaced
  with explicit stack manipulation operations.  $\mathrm{Push}$ and $\mathrm{Pop}$
  save and load data; $\mathrm{PushJump}\ i\ j$ jumps to block $i$ after setting up
  a return to block $j$; and $\mathrm{Return}$ returns by popping the program counter stack.}
\label{fig:tensor-stack-syntax}
\end{figure}

\begin{algorithm}[htp!]
  \caption{Program counter autobatching}
  \label{alg:tensor-stack}
\begin{algorithmic}
   \STATE {\bfseries Input:} Program $P$ with $I$ basic blocks $B_i$, input variable $x$, and output variable $y$;
   \STATE {\bfseries Input:} Batch size $Z$; Stack depth limit $D$;
   \STATE {\bfseries Input:} Data array $T$ with leading dimension $Z$.
   \STATE Initialize $D$-by-$Z$ program counter $pc = [0, 0, \ldots, 0]$
   \STATE Initialize length $Z$ stack indexes $pc_{stack} = [0, 0, \ldots, 0]$
   \FOR{variable $v$}
   \STATE Initialize $v$ to zeros with leading dimensions $D, Z$
   \STATE Initialize length $Z$ indexes $v_{stack} = [0, 0, \ldots, 0]$
   \ENDFOR
   \STATE PUSH $T$ onto $x$
   \STATE Initialize length $Z$ $pc^{top} = pc[pc_{stack}]$
   \WHILE{any $pc^{top} < I$}
   \STATE Set next block index $i = \min{pc^{top}}$
   \STATE Compute locally active set $A = \{b | pc^{top}_b = i\}$
   \FOR{$op \in B_i$}
   \IF{$op$ is $\mathrm{Push}\ y = f(x)$}
   \STATE Compute $x^{top} = x[x_{stack}]$
   \STATE Compute outputs $o = f(x^{top})$
   \STATE PUSH $o_{A}$ onto $y_{A}$
   \ELSIF{$op$ is $\mathrm{Pop}\ x$}
   \STATE POP $x_{A}$
   \ENDIF
   \ENDFOR
   \IF{$t_i$ is $\mathrm{Jump}\ j$}
   \STATE Set $pc^{top}_{A} = j$
   \ELSIF{$t_i$ is $\mathrm{Branch}\ x\ j\ k$}
   \STATE Compute $x^{top} = x[x_{stack}]$
   \FOR{$b \in A$}
   \STATE Set $pc^{top}_b = j$ if $x^{top}_b$ otherwise $pc^{top}_b = k$
   \ENDFOR
   \ELSIF{$t_i$ is $\mathrm{PushJump}\ j\ k$}
   \STATE Set $pc^{top}_{A} = k$
   \STATE PUSH $j$ onto $pc_A$
   \ELSIF{$t_i$ is $\mathrm{Return}$}
   \STATE POP $pc_A$
   \ENDIF
   \ENDWHILE
   \STATE \textbf{return} Current value of $y[y_{stack}]$
\end{algorithmic}
\end{algorithm}

To implement this, we want a slightly different control flow graph
language, shown in Figure~\ref{fig:tensor-stack-syntax}.  Since the
runtime is now managing the stacks itself, we replace the
$\mathrm{Call}$ instruction with explicit (per-variable)
$\mathrm{Push}$ and $\mathrm{Pop}$ instructions, as well as
$\mathrm{PushJump}$ for entering function bodies.  The $\mathrm{Push}$ also computes
the value to write to the top of the variable's stack.  The language is
otherwise the same as Figure~\ref{fig:python-stack-syntax}, and indeed
our implementation compiles to the latter first and then lowers from
there to the former.

\subsection{Runtime}

The runtime is spelled out in Algorithm~\ref{alg:tensor-stack}.  As
compared with local static autobatching
(Algorithm~\ref{alg:python-stack}), the active set no longer persists
across steps, while the explicit program counter takes on a more
central role (hence the name).  The program counter now has a stack
dimension of its own.  The locally active set can now include batch
members at different stack depths, giving the runtime a chance to
batch their computations together.  Consequently, computations
can converge by calling into the same subroutine from different
code locations, and conversely diverge by returning thereto.

The major advantage of managing variable stacks and the program
counter is that the runtime is no longer itself recursive, so can be
coded completely in a system like (graph-mode) TensorFlow or XLA that
does not support recursion natively.  In TensorFlow, the code
corresponding to the main loop in Algorithm~\ref{alg:tensor-stack}
looks like Figure~\ref{fig:pseudocode}.  The loop body is a dispatch
to the correct block based on the dynamic value of the next program
counter.  The main trick, common in partial evaluation
\cite{jones1993partial}, is that the block index \texttt{pc} on line 3
is static at TF graph build time.  The \texttt{block} function (not
shown) is hence just a direct transcription of the loop body in
Algorithm~\ref{alg:tensor-stack}.  Picking out the \texttt{pc}th
program block and interpreting that block's operations will happen
once when constructing the TensorFlow graph, leaving just the Tensor
operations needed to manipulate the autobatched variables.  In effect,
lines 3--4 are the final stage of our compiler, lowering the entire
autobatched program from our representation of
Figure~\ref{fig:tensor-stack-syntax} to raw TensorFlow operations.

\begin{figure}[thp]
\begin{minted}[
    mathescape,
    linenos,
    numbersep=5pt,
    frame=none,
    framesep=2mm]{python}
while next_pc < halt_pc:
  all_vars = tf.switch_case(next_pc,
    [block(pc, all_vars)
     for pc in valid_pcs])
  next_pc = tf.reduce_min(
    get_top(get_pc_var(all_vars)))
\end{minted}
\caption{Pseudocode for implementing the main loop of
  Algorithm~\ref{alg:tensor-stack} in TensorFlow.
  The main idea is to dispatch at runtime on the dynamic $\mathtt{next\_pc}$ (line 2),
  to the block compiled with the correct static $\mathtt{pc}$ (line 3).  See text.}
\label{fig:pseudocode}
\end{figure}

\subsection{Optimizations}

The price we pay for implementing our own stack is that reads from
the stack structure must gather according to the stack depths (which
may vary across batch members), and writes must correspondingly
scatter.  We implement five compiler optimizations to reduce
this cost:
\begin{enumerate}
\item Each variable in the program being auto-batched gets its own
  stack, so we can optimize those stacks independently.  In
  particular, we can arrange the stack operations in a per-variable
  caller-saves stack discipline to set up for later pop-push
  elimination.
\item The compiler statically detects whether each variable is live
  across an iteration of the runtime loop.  Those that are not are
  temporary and don't need to be touched by the autobatching system at
  all: they exist only inside basic block executions.
\item The compiler also statically detects whether each variable is
  live across a recursive function call that might need to reuse it
  (at a different stack depth).  Those that are not do not need a
  stack or a stack pointer, and autobatching only amounts to updating
  their top value with a mask to select only the active batch members.
\item For those variables that do require stacks, the runtime
  caches the top of each stack variable, so that repeated
  reads do not require large gather operations.
\item Finally, the compiler also statically cancels pairs of $\mathrm{Pop}$
  followed by $\mathrm{Push}$ that have no intervening reads, and
  converts the latter into an in-place $\mathrm{Update}$ instruction (not shown)
  that only interacts with the cached top of each stack.
\end{enumerate}

An important consequence of the above optimizations is that program
counter autobatching will run a non-recursive program entirely without
variable stacks (except for the program counter itself).  It will thus
replicate the performance of local static autobatching without needing
a host-language stack for (non-recursive) function calls, while also
being able to batch across them.  Unlike a tracing-based system such as JAX \cite{jax}, this
compiled approach also doesn't amount to inlining all function calls,
so can autobatch a program with significant subroutine reuse without
combinatorial explosion in code (or traced graph) size.

\begin{figure*}[ht]
  \includegraphics{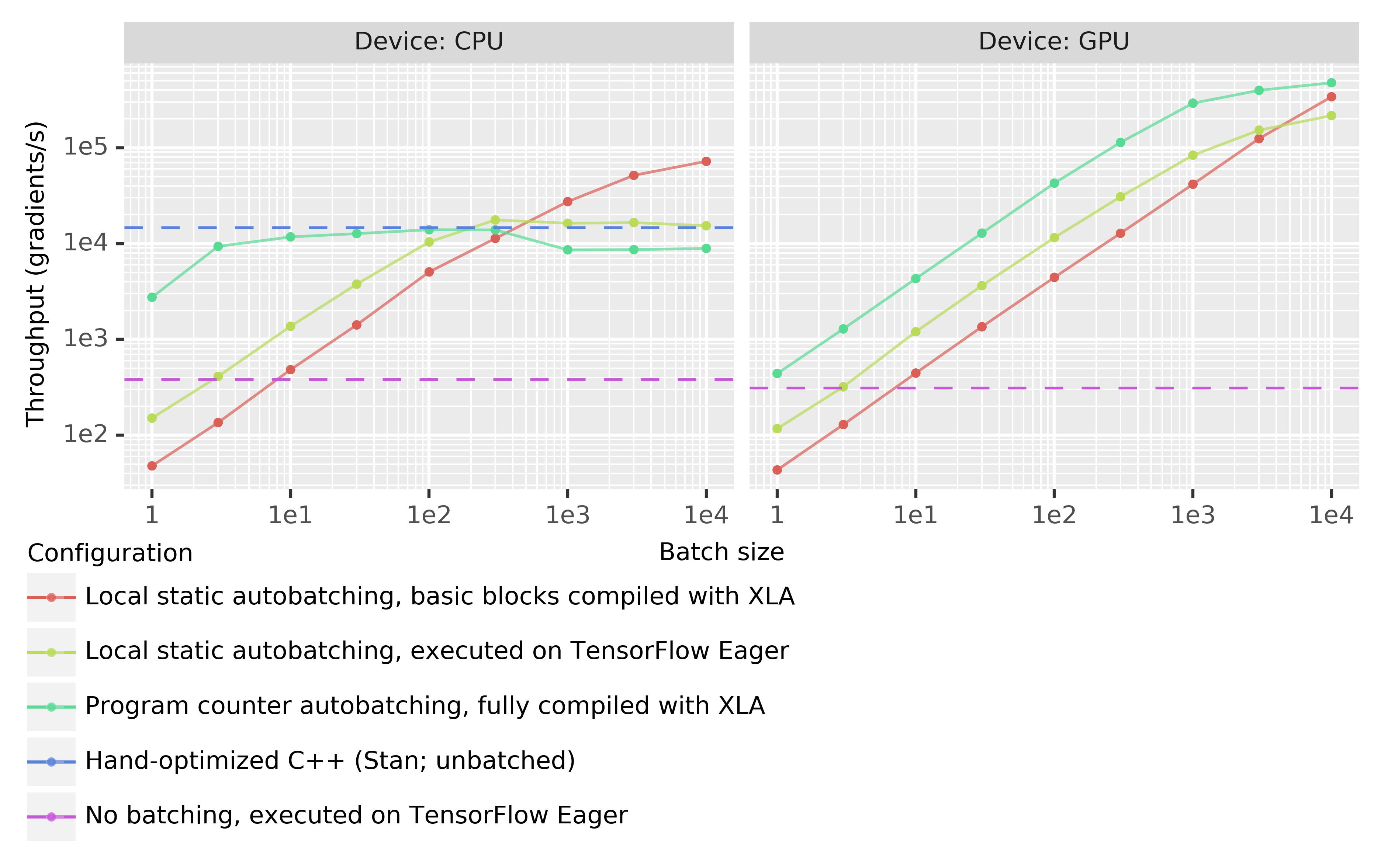}
  \caption{Performance of auto-batched No U-Turn Sampler on the
    Bayesian logistic regression problem (100
    latent dimensions, 10,000 data points).  The batch
    size refers to the number of chains running in tandem.  The
    reported gradients are the total across all chains, excluding
    waste due to synchronization.  We compare
    the performance of program counter autobatching compiled with XLA
    to our local static autobatching executed in TensorFlow's Eager
    mode.  We also include two baselines.  One is the same program
    executed directly in Eager mode without autobatching (perforce
    running one batch member at a time).  The other is the widely used
    and well-optimized Stan implementation of (a variant of) the same
    NUTS algorithm.  Batching provides linear scaling on all tested
    platforms, until the underlying hardware saturates.  See text for
    details of the experimental setup.}
  \label{fig:nuts-results}
\end{figure*}

\section{Experiments}
\label{sec:results}

We evaluate autobatching on the No U-Turn Sampler (NUTS), a workhorse
gradient-based Markov-chain Monte Carlo method widely used in Bayesian
statistics.  We chose NUTS as a test target because (i) the standard
presentation is a complex recursive function, prohibitively difficult
to batch by hand, and (ii) batching across multiple independent chains
can give substantial speedups.

Our results show that batching on GPUs enables NUTS to efficiently
scale to thousands of parallel chains, and get inferential throughput
well beyond existing CPU-based systems such as Stan.  By demonstrating performance gains
from batching NUTS, we hope to contribute to a broader practice of
running large numbers of independent Markov chains, for more precise
convergence diagnostics and uncertainty estimates.

We test autobatched NUTS on two test problems:
\begin{itemize}
\item Exploring a 100-dimensional correlated Gaussian distribution.
\item Inference in a Bayesian logistic regression problem with 10,000
  synthetic data points and 100 regressors.
\end{itemize}

We test three forms of autobatching NUTS:
\begin{itemize}
\item Program counter autobatching, compiled entirely with XLA;
\item Local static autobatching, executed entirely with TensorFlow Eager; and
\item A hybrid: Running the control operations of local static
  autobatching in TensorFlow Eager, but compiling the straight-line
  components (basic blocks) with XLA.
\end{itemize}
The purpose of the latter is to try and tease apart the benefits of
compilation specifically for control flow versus compiling (and fusing
together) straightline sequences of kernel invocations.  It should be
noted that identifying the basic blocks to compile separately is a
nontrivial program transformation in its own right.  It fits
conveniently into our software framework, but represents considerable
work to do by hand.

\subsection{Runtime on logistic regression}

In Figure~\ref{fig:nuts-results}, we measure the number of model
gradient evaluations per second we can get out of batching NUTS in
different ways.  The main effect we see is GPU throughput scaling
linearly with batch size.  We also see the speedup from avoiding cycling
to Python (on the host CPU!) to implement the recursion.

The behavior when running entirely on CPU is more nuanced.  CPUs are
superb at control flow and recursion as it is, so the main effect of
batching seems to be to amortize away platform overhead, until we
match the performance of the Stan system's long-optimized custom C++ at
a batch size of a few hundred---or just ten for compiling fully with
XLA, whose per-leaf-kernel overhead is much smaller.

At very large batch sizes, however, the hybrid strategy of running
local static autobatching in TensorFlow Eager but compiling the basic
blocks with XLA outperforms all other NUTS implementations we tested.
We are not quite sure why
this happens, but we hypothesize that (1) it beats Stan because of
better memory locality in batch evaluation of the target density; (2)
it beats fully compiled autobatching by avoiding the overhead of
gathering from and scattering to per-variable stack arrays; (3) it
beats fully Eager local autobatching by avoiding per-leaf-kernel
dispatch overheads; but (4) it's slower at low batch sizes because of
per-fused-kernel invocation costs.  We leave a complete investigation of
this phenomenon to future work.

\begin{figure*}[ht]
  \includegraphics{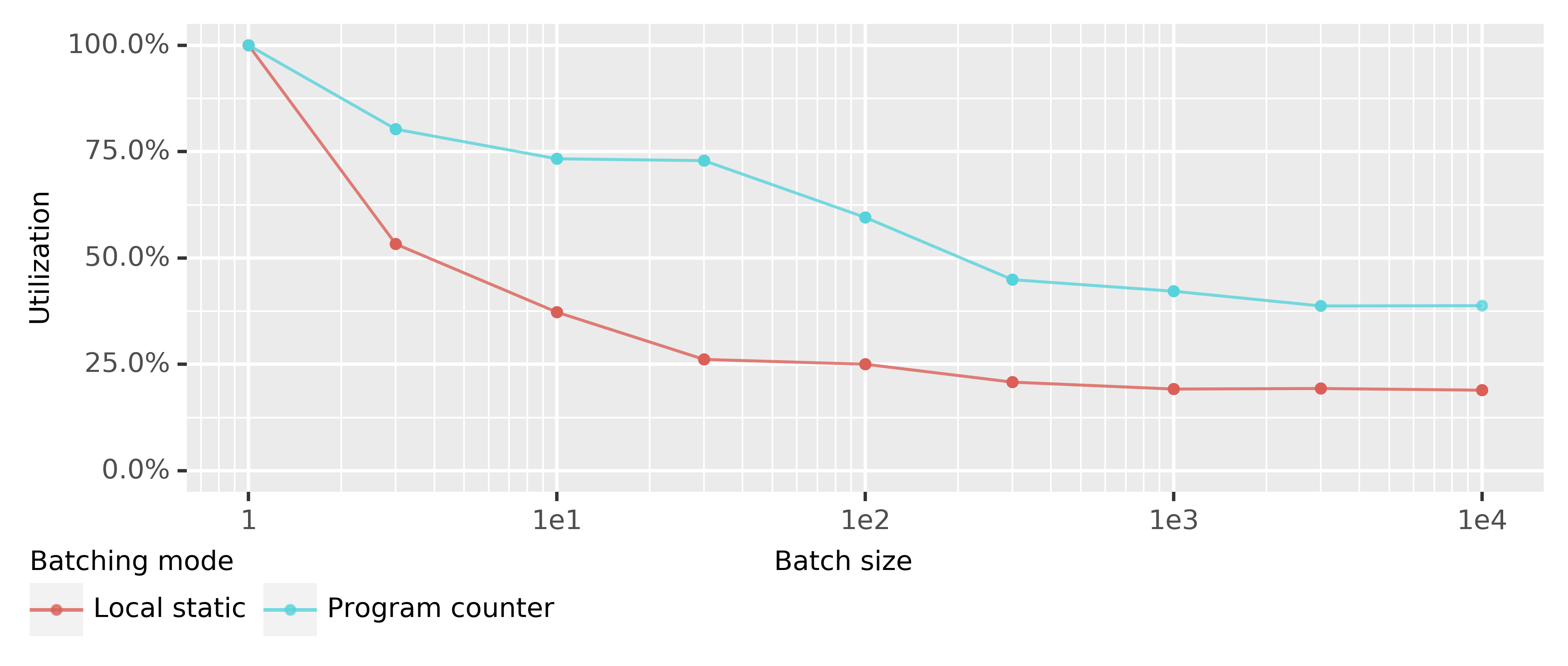}
  \caption{Utilization of batch gradient computation on the correlated
    Gaussian test problem.  Utilization is less than 100\% above 1
    batch member because different batch members choose to use
    different numbers of gradients at each trajectory.  We can see
    from the local-static line that on this problem, the longest
    trajectory that NUTS chooses at any iteration tends to be about
    four times longer than the average.  Program counter autobatching
    recovers more utilization by batching gradients across 10 consecutive
    NUTS trajectories, instead of having to synchronize on trajectory
    boundaries.}
  \label{fig:nuts-utilization}
\end{figure*}

A few details of the experimental setup.  These measurements are for
the synthetic Bayesian logistic regression problem.  The measured time
counts only a warm run, excluding compilation, the one-time TensorFlow
graph construction, etc.  This allows measurements for small batch
sizes not to be swamped by O(1) overhead.  The CPU computations were
run on a shared 88-core machine with 100 GiB of RAM allocated to the
benchmark, in 32-bit floating-point precision.  The GPU computations
were run on a dedicated Tesla P100 GPU, also in 32-bit precision.  In
all cases, we slightly modified the published NUTS algorithm to take 4
steps of the leapfrog integrator at each leaf of the NUTS tree, to
better amortize the control overhead.  This has no effect on the
soundness of the algorithm.  The timings are best of five independent
runs.  Due to technical limitations, the Stan baseline was run on
different hardware.  We scaled its throughput against a calibration
run of program counter autobatching on the same machine and precision.

\subsection{Batch utilization on correlated Gaussian}

We also measure the specific effect of batching across recursion
depths.  The NUTS algorithm dynamically
chooses how many gradient steps to take in each trajectory.  When
running a multi-step Markov chain, one therefore has a choice of
whether to synchronize on trajectory boundaries or on individual
gradient steps.  Local autobatching systems can only implement the
former, because the control structure of the whole batch computation
necessarily follows the control structure of the user program.
Program counter autobatching, however, can synchronize on the
gradients, for example evaluating the 5th gradient of the 3rd trajectory of one
batch member in tandem with the 8th gradient of the 2nd trajectory of
another.  In the regime where the model gradients are expensive
relative to the inter-trajectory book keeping, the latter should be
preferable.

We find in Figure~\ref{fig:nuts-utilization} that on a synthetic
correlated Gaussian, synchronizing on trajectory boundaries leaves as
much as a factor of 4 on the table even at a batch size as low as 30.
Program counter autobatching is able to recover a factor of about 2 in
utilization in as few as 10 NUTS trajectories.  We expect gradient
utilization to approach 1 in the limit of long chains as the per-chain
distribution of total gradients taken approaches a Gaussian.


\section{Related Work}
\label{sec:related-work}

The machine learning community has seen several systems arise to
address the difficulty of hand-batching by batching user programs
automatically.  The extant ones have used one of two major
architectures.  The first is the local static autobatching we
described in Section~\ref{sec:python-stack}.  The second, called
\emph{dynamic batching}, is exemplified by \cite{neubig2017fly} and
\cite{looks2017deep}.  In this architecture, the runtime performs
batching dynamically, by running parallel evaluations of the user
program against a scheduler that manages the execution and batches
opportunistically.  From the lens of control flow, the advantage of
dynamic batching is its ability to recover more batching (including
within a single execution, if there is no data dependence).
On the other hand, both local and program counter autobatching have less runtime
overhead, because all the decisions about batch scheduling are done
statically (at batch-program extraction time).  For the same reason,
the latter two architectures are more amenable to running on top
of an existing machine learning framework, whereas dynamic batching
requires support from the underlying graph executor.

The presentation in Section~\ref{sec:python-stack} gives one
implementation style for the local static autobatching transformation.
Other systems achieve the same effect in different ways.

Matchbox \cite{matchbox-systems-for-ml-workshop-2018}
relies on a relatively lightweight syntax
transformation to intercept \texttt{if} statements and \texttt{while}
loops, and otherwise accomplishes batching by defining a ``batched
array'' type that carries the mask.  The batched array overloads all
the methods for a standard array with appropriate additional masking.
Recursion is left to the ambient Python stack.  In our terms, the mask
corresponds to the active set.

At \texttt{if} statements, Matchbox first executes the \texttt{then}
arm (if any batch members need it) and then the \texttt{else}.  The
program counter of Algorithm~\ref{alg:python-stack} is thus encoded in
the queue (also maintained on the Python stack) of mask-block pairs to
be executed.  The data structure is equivalent: one vector of indices
encodes the same information as a list of pairs of index with
exclusive mask of items having that index.  Storing the queue of
program resumption points on the Python stack makes it more
difficult for Matchbox to use a different heuristic for the order in
which to run blocks, but the extant behavior is equivalent to the
``run the earliest block'' heuristic presented in
Section~\ref{sec:python-stack}.

Jax \cite{jax} relies on an explicit tracing pass to construct
an internal representation, on which batching (invoked via
\texttt{jax.vmap}) is an explicit static transformation (one of
several Jax can perform).  Control flow requires using the
Jax-specific control operators: \texttt{lax.cond} instead of
\texttt{if} and \texttt{lax.while\_loop} instead of \texttt{while}.

Recursion is not supported in Jax at all, because it confuses the
tracer.  There is therefore no stack.  The program counter is encoded
in a mask and an execution sequence the same way it is in Matchbox,
with the same effects.

Similarly, TensorFlow's \texttt{pfor} facility \cite{pfor-icml-2019, agarwal2019autovectorizing} operates on
the TensorFlow graph, including its \texttt{tf.cond} and
\texttt{tf.while\_loop} control operators.  The transformation in
\texttt{pfor} is the same as Jax's \texttt{vmap}, up to
implementation details.  Recursion is similarly not supported, because
the underlying TensorFlow graph doesn't support it.

This is the sense in which the transformation is ``local'': this
autobatching style (at least with this basic block choice heuristic)
perserves the nesting structure of the user's original control
constructs.  As such, it can be implemented by a local transformation
that, e.g., turns a \texttt{while} loop into a similar loop with a
transformed body.

Somewhat farther afield, GPU programming languages such as CUDA \cite{Nickolls:2008:SPP:1365490.1365500}
are also automatically vectorized.  The handling of
control constructs in CUDA is identical with local static
autobatching, but
of course only applies to kernels written therein.  An interesting
potential direction for application-level automatic batching could be
to forward surface language control constructs through a compiler
targeting CUDA (such as the GPU backend of the XLA compiler) and rely
on CUDA to batch them.  The ISPC compiler \cite{pharr2012ispc} performs
the same automatic vectorization transform for vector units in CPUs.

Finally, the NUTS algorithm is central enough to have prompted two rewrites
in non-recursive form \cite{iterative-nuts, unrolled-nuts} for the express purpose
of running it on accelerators more effectively.  The non-recursive
form is also amenable to batching either by hand or using an
established autobatching system (whether local or dynamic).  One would
expect such a manual effort to obtain better performance, but its
labor-intensiveness necessarily limits its scope.

\section{Conclusion}
\label{sec:conclusion}

We presented program counter autobatching, a novel method for
automatically vectorizing batch computations at the machine learning
framework level.  Program counter autobatching handles arbitrary
control flow in the source program, including batching operations
across recursion depth.  We demonstrated the efficacy of the method by
mechanically batching a (recursive) implementation of the No U-Turn
Sampler, obtaining speedups varying (with batch size) up to
three orders of magnitude.  An implementation of program counter
autobatching is available in the TensorFlow Probability package.

\section*{Acknowledgments}

The authors would like to thank Delesley Hutchins for invaluable early
critique of the architecture of the compiler; Sean Talts for
benchmarking against Stan; Pavel Sountsov, Alex Wiltschko, James
Bradbury, and Dan Piponi for insightful discussions; and the anonymous
reviewers for their feedback.

\bibliography{main}
\bibliographystyle{mlsys2020style/mlsys2020}

\appendix

\end{document}